\documentclass[aps,prl,twocolumn,superscriptaddress,showpacs]{revtex4}
\usepackage{graphicx}

\newlength{\intwidth}

\def\lapprox{\,\raise0.4ex\hbox{$<$}\kern-0.8em\lower0.7ex\hbox{$\sim$}\,}
\def\gapprox{\,\raise0.4ex\hbox{$>$}\kern-0.8em\lower0.7ex\hbox{$\sim$}\,}
\def\lg{\,\raise0.5ex\hbox{\footnotesize $<$}\kern-0.8em\lower0.5ex\hbox{\footnotesize  $>$}\,}
\def\gl{\,\raise0.5ex\hbox{\footnotesize $>$}\kern-0.8em\lower0.5ex\hbox{\footnotesize  $<$}\,}

\def\be{\begin{equation}}
\def\ee{\end{equation}}
\def\ba{\begin{eqnarray}}
\def\ea{\end{eqnarray}}

\def\bc{\begin{center}}
\def\ec{\end{center}}

\begin{document}

\title{Cyclotron spin-flip excitations in a $\nu\!=\!1/3$ quantum Hall ferromagnet}

\author{A.~B.~Van'kov}
\affiliation{Institute of Solid State Physics, RAS, Chernogolovka,
142432 Russia}

\author{L.~V.~Kulik}
\affiliation{Institute of Solid State Physics, RAS, Chernogolovka,
142432 Russia}

\author{S.~M.~Dickmann}
\affiliation{Institute of Solid State Physics, RAS, Chernogolovka,
142432 Russia}

\author{I.~V.~Kukushkin}
\affiliation{Institute of Solid State Physics, RAS, Chernogolovka,
142432 Russia} \affiliation{Max-Planck-Institut f\"ur
Festk\"orperforschung, Heisenbergstr. 1, 70569 Stuttgart, Germany}

\author{V.~E.~Kirpichev} \affiliation{Institute of Solid State
Physics, RAS, Chernogolovka, 142432 Russia}

\author{W.~Dietsche}
\affiliation{Max-Planck-Institut f\"ur Festk\"orperforschung,
Heisenbergstr. 1, 70569 Stuttgart, Germany}

\author{S.~Schmult}
\affiliation{Max-Planck-Institut f\"ur Festk\"orperforschung,
Heisenbergstr. 1, 70569 Stuttgart, Germany}

\date{\today}

\begin{abstract}
Inelastic light scattering spectroscopy around the $\nu\!=\!1/3$ filling discloses a novel type
of cyclotron spin-flip excitation in a quantum Hall system in addition to the excitations
previously studied. The excitation energy of the observed mode follows qualitatively the degree
of electron spin polarization, reaching a maximum value at $\nu\!=\!1/3$ and thus characterizing
it as a $\nu\!=\!1/3$ ferromagnet eigenmode. Its absolute energy substantially exceeds the
theoretical prediction obtained within the renowned single-mode approximation. Double-exciton
corrections neglected utilizing the single-mode approach are evaluated within the framework of
the excitonic representation and are inferred to be responsible for the observed effect.

\end{abstract}


\maketitle

Physics of two-dimensional (2D) electron systems in a strong perpendicular magnetic field is
governed by the macroscopic degeneracy of electron states in Landau levels. Owing to this point
and because of the strong many-particle interaction the ground state at the unit filling of the
zeroth Landau level (LL) is an itinerant ferromagnet where electron spins tend to align even for
arbitrarily small Zeeman coupling. The ground  state at fractional filling $\nu\!=\!1/3$ is
believed to be somewhat analogous. In fact, the magnetization around both $\nu\!=\!1/3$ and
$\nu\!=\!1$ behaves similarly. It reaches pronounced maxima at those particular filling factors
\cite{Barrett1_3,Barrett1}, which is generally interpreted in favor of formation of the quantum
Hall ferromagnet.

A natural way to characterize a ferromagnetic state is to study its spin-flip excitations and
their modification under the influence of external parameters (magnetic field, temperature,
etc.). Most relevant for the $\nu\!=\!1$ quantum Hall ferromagnet are the spin exciton (spin
waves) and the cyclotron spin-flip excitation (CSFE) modes. Experimentally accessible
long-wavelength spin excitons carry little information about the ferromagnetic order by virtue
of the Larmor theorem \cite{Larmor}. The major efforts have therefore been concentrated hitherto
around the CSFE
--- a long-wavelength excitation simultaneously changing the orbital and spin quantum numbers of
a 2D electron gas ($\delta n\!=\!1$ and $\delta S\!=\!\delta S_z\!=\!-1$). Studies of the CSFE
by means of the inelastic light scattering spectroscopy revealed peculiarities of the Coulomb
interaction in two dimensions, disclosed new magnetic phases, and helped to draw qualitative
conclusions about thermodynamics of the quantum Hall ferromagnet \cite{nu1our,nu1temp}.
Surprisingly, similar experiments at $\nu\!=\!1/3$ did not produce such clear results. The
cyclotron spin-flip mode was detected, yet its energy reflected rather the total electron
density in the 2D electron gas (2DEG) than a variation of the electron magnetization
\cite{Kulik2001}. On the other hand if the electron magnetization has a pronounced maximum at
$\nu\!=\!1/3$ so should the CSFE energy as a measure of the ground state exchange interaction.
The apparent contradiction between naive expectations and the experiment is resolved in the
present work where we report on a new cyclotron spin-flip mode whose energy variation matches
the electron magnetization. The energy of this mode is evaluated within the excitonic
representation, and the problem of why two cyclotron spin-flip modes with different energies
coexist in the available 2D electron system is discussed.

Two high-quality heterostructures were studied. Each consisted of a single-side doped 25\,nm
GaAs/Al$_{0.3}$Ga$_{0.7}$As quantum wells (QWs) with electron densities of 1.2 and $2.2 \times
10^{11}$\,cm$^{-2}$ in the dark. The mobilities were 7 and $10 \times 10^{6}$\,cm$^2$/V$\cdot$s.
The electron densities were tuned via the opto-depletion effect and were measured by means of
in-situ photoluminescence \cite{Kukushkin96}. The inelastic light scattering experiment was
performed in the temperature range of $0.3\div 1.5$\,K in the magnetic field up to 14\,T normal
to the sample surface. Inelastic light scattering spectra were obtained using a Ti:sapphire
laser tunable above the fundamental band gap of the QW. The laser excitation energy was varied
between 1.545 and 1.575\,eV. The power density was below $0.1$\,W/cm$^2$. A two-fiber optical
system was employed in the experiments \cite{2fiber}. One fiber transmitted the pumping laser
beam to the sample, the second fiber collected the scattered light and guided it out of the
cryostat. The scattered light was dispersed either by a T-64000 triple spectrograph or a {
Ramanor} U-1000 double-grating monochromator and recorded with a charge-coupled device camera.
The spectral resolution of the overall detection system was about 0.03\,meV.

Figure\,1 presents the key experimental result. Inelastic light scattering lines corresponding
to two cyclotron spin-flip modes are clearly seen above the cyclotron energy. One (A) coincides
with the ILS line found earlier in terms of energy and linewidth \cite{Kulik2001}. The energy of
the line A increases monotonically with the electron filling factor not showing any
peculiarities around $\nu\!=\!1/3$. In contrast to this, the second line (B) demonstrates a
non-monotonic energy dependence with a pronounced maximum at $\nu\!=\!1/3$. Moreover, the energy
of line B is proportional to the 2DEG magnetization measured by Khandelwal {\it et al}. under
similar experimental conditions \cite{Barrett1_3}. This links the line B to the inelastic light
scattering on a spin-flip mode of the $\nu\!=\!1/3$ quantum Hall ferromagnet. Yet, the question
arises how two cyclotron spin-flip modes with so different properties could possibly exist in a
2DEG.

\begin{figure}[htb!]
\includegraphics[width=.48\textwidth]{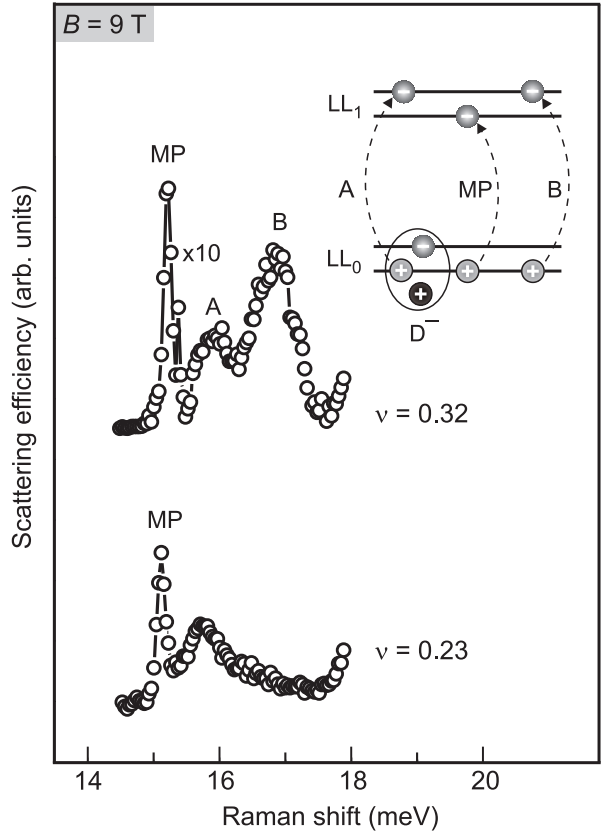}
\caption{\label{fig3} Inelastic light scattering spectra at $B\!=\!9\,$T and two different
filling factors\,-- $\nu\!=\!0.33$ and $\nu\!=\!0.23$. Line B is the cyclotron spin-flip
excitation (CSFE) of the $\nu\!=\!1/3$ quantum Hall ferromagnet. Line A corresponds to the
analogous excitation of $D^-$ complexes. At filling factors far from $\nu\!=\!1/3$ line B
disappears from the spectrum (see the lower spectrum). The inset schematically explains the
origin of the observed lines.}
\end{figure}

Several aspects of this problem have already been clarified in our previous publications. It was
shown \cite{nu1temp,Donors} that under certain experimental conditions a 2D electron system in
the vicinity of a positive charge is unstable against formation of a spin-singlet barrier $D^-$
complex -- two electrons in a QW with oppositely aligned spins bound to an impurity in the QW
barrier. In the quantum limit the barrier $D^-$ complexes occupy a significant part of the
experimentally accessible 2DEG even in the highest quality AlGaAs/GaAs QWs. Because of two
electron subsystems ($D^-$ complexes and unbound electrons), two spin-flip modes exist. The line
A arises from the upper branch of spin-flip excitations of barrier $D^-$ complexes (see the
inset in Figure~1), whereas the line B originates from the free electron gas. Note, that the two
subsystems are not really independent. The electrons of $D^-$ complexes interact with the
unbound electrons, i.e. they are in fact many-particle conglomerates which become truly isolated
only in the extreme quantum limit $\nu\!<\!1/10\,$ \cite{Kulik2001}. In other words, the $D^-$
complexes supply a degree of freedom to form an additional spin-flip mode, whose energy is not
sensitive to electron magnetization. We do not consider the line A carefully studied in Ref.
[\onlinecite{Kulik2001}] and focus on the line B as an eigen CSFE mode of the clean quantum Hall
ferromagnet.

\begin{figure}[htb!]
\includegraphics[width=.48\textwidth]{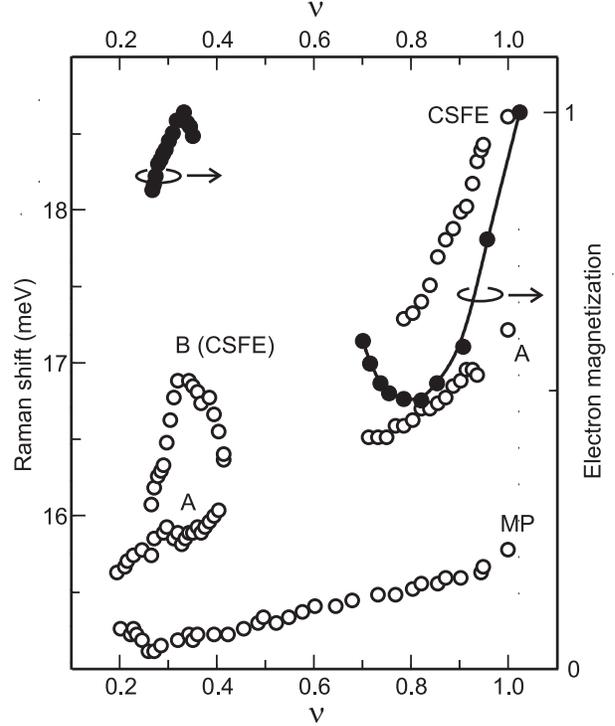}
\caption{\label{fig4} Experimental $\nu$ dependence of the energies for lines A and B around
$\nu\!=\!1/3$. Also shown are the magnetoplasmon energy (MP) and energies for the corresponding
lines in the vicinity of unit filling (at $\nu\!<\!1$). Black points show electron magnetization
measured with optically detected magnetic nuclear resonance in Ref.
[\onlinecite{Barrett1,Barrett1_3}] under experimental conditions similar to ours.}
\end{figure}

The CSFE energy behaves much alike at $\nu\!=\!1/3$ and $\nu\!=\!1$ (Figure\,2). However, unlike
the $\nu\!=\!1$ ferromagnet, the $\nu\!=\!1/3$ state is highly correlated. The CSFE energy at
$\nu\!=\!1/3$ is therefore determined not only by the exchange interaction, as it is at
$\nu\!=\!1$, but also by electron-electron correlations in the ground and excited states. To
shed a light on the interplay between exchange and correlations in the $\nu\!=\!1/3$
ferromagnet, the experimental CSFE energy is compared with the existing theories, and a new
theoretical approach to the calculation of the CSFE energy is developed. First of all we remind
the reader that even in the $\nu\!=\!1$ ferromagnet the CSFE is not a single-mode
(single-exciton) excitation but has a double-exciton component consisting of a spin wave (with
the change of quantum numbers: $\delta S\!=\!\delta S_z\!=\!-1$, and $\delta n\!=\!0$) and a
magnetoplasmon (with $\delta S\!=\!\delta S_z\!=\!0$, and $\delta n\!=\!1$) \cite{di08}.
However, in the renown single mode approximation (SMA) the double-exciton component is ignored.
Besides, if the Hartree-Fock (HF) approximation is used to describe the ground state (GS), the
SMA result for the CSFE correlation shift is easily generalized to arbitrary fillings
$\nu\!<\!1$, and at zero momentum it reads as follows:
$$
  {\cal E}_{\rm SM}^{\rm HF}
 =\nu\frac{e^2}{2\kappa l_B}\int_0^{\infty}{p^3dp}V(p)e^{-p^2/2}  \eqno (1)
 $$
(this energy is counted from the cyclotron energy $\hbar\omega_c$) \cite{Pinczuk1}. Here $l_B$
is the magnetic length and $2\pi V(q)$ is the dimensionless Fourier component of the effective
{\it e-e} interaction vertex in the 2D layer. (In the ideal 2D case $V(q)\!=\!1/q$ but actually
$V(q)\!=\!F(q)/q$, where $F(q)$ is the geometric form-factor; the ratio $r_{\rm
c}\!\!=\!(e^2\langle F\rangle/\kappa l_B)/\hbar\omega_c$, where $\langle F\rangle$ is the
averaged value, is considered to be small.) Calculations with formula (1) employing the usual
self-consistent procedure to find the form-factor $F(q)$ \cite{lu93}, yield in our specific case
($B\!=\!9\,$T, $\nu\!=\!1/3$)  ${\cal E}_{\rm SM}^{\rm (HF)}\!\!=\!1.26\,$meV. This is rather
far off the experimental value ${E}_{\rm exp}\!\!=\!1.69\,$meV observed in the case (see
Fig.~2).

A more refined approach utilizes Laughlin's wave function for the ground state to calculate the
CSFE energy within the SMA approach \cite{lo93}. The SMA result for the correlation shift would
be ${\cal E}_{\rm SM}\!=\!0.79\,$meV in our case. Being even smaller than the HF result (1), it
is in a striking disagreement with the experiment. Another development of the existing theory
should account for the multi-component feature of CSFE. In a fractional quantum Hall
ferromagnet, charge-density waves (intra Landau level excitations) are split from the GS by a
considerable energy gap (see, e.g., Ref. [\onlinecite{gi86}]). Due to this fact, we will use a
model where charge-density waves (CDWs) are ignored but the double-exciton component
corresponding to coupled spin wave and magnetoplasmon is taken into account. Our present
approach is thus a projection of the $\nu\!=\!1$ CSFE theory \cite{di08} onto the fractional
$\nu$ case. Specifically, to estimate the CSFE energy we use the double-mode approximation
(DMA). The CSFE state is represented as
$$
|{\rm SF}\rangle\!=\!{\cal Q}^\dag_{0\overline{1}{\bf 0}}|0\rangle\!+\!AN_\phi^{-1/2}\sum_{\bf
q}\psi(q){\cal Q}_{0\overline{0}-\!{\bf q}}^\dag {\cal Q}_{01{\bf q}}^\dag|0\rangle  \eqno (2)
$$
($N_\phi$ is the magnetic flux number in the 2DEG studied). The definition of the exciton
creation operator ${\cal Q}_{ab{\bf q}}^\dag$ in terms of electron annihilation and creation
operators can be found e.g., in Ref. [\onlinecite{di08}] (see also commutation rules for the
${\cal Q}$-operators there, and references therein). Here $a$ and $b$ are binary indexes
labelling LLs and spin sublevels: $a\!=\!0$ to denote the $(n,S_z)\!=\!(0,\uparrow)$ state,
$b\!=\!\overline{0}$ is the $(0,\downarrow)$ state, $b\!=\!{1}$ corresponds to
$(n,S_z)\!=\!(1,\uparrow)$ and $b\!=\!\overline{1}$ is the $(1,\downarrow)$ state. In the
$\nu\!\leq\!1$ case we have $\langle 0|{\cal Q}_{ab{\bf q}}{\cal Q}_{ab{\bf
q}'}^\dag|0\rangle\!=\!\nu\delta_{a,\,0}\delta_{{\bf q},\,{\bf q}'}$ if $b\!\not=\!0$. We
consider the ground state $|0\rangle$ in the HF approximation. Then one can also find that
$\langle 0|{\cal Q}_{aa{\bf q}}{\cal Q}_{aa{\bf
q}'}^\dag|0\rangle\!=\!\nu\delta_{a\!,0}\delta_{{\bf q},\,{\bf q}'}\!\left(1\!-\!\nu\!+\!\nu
N_\phi\delta_{{\bf q},\,0}\right)$; c.f. Ref. [\onlinecite{di08}].

Due to the same reasons as those presented in Ref. [\onlinecite{di08}], the ``wave function''
$\psi(q)$ in Eq. (2) is chosen to be equal to $L_1(q^2)e^{-q^2/2}$. ($L_1$ is the Laguerre
polynomial.) Similar to the $\nu\!=\!1$ case, after variational procedure we find the fitting
parameter $A$ and the CSFE correlation shift. The latter again is  the largest root of a
$2\times 2$ secular equation. Now it takes the form \vspace{1mm}
$$\vspace{1mm}
{}\!{}\!{}\!{}\!{}\!{}\mbox{det}\!\left|(E\!-\!{\cal
E}_i)\delta_{i\!,\,k}\!+\!(1\!-\!\delta_{i\!,k}){\cal D}_{ik}\right|\!=\!0\;\; (i,k=1\;\,
\mbox{or}\; 2),\! \eqno (3)
$$
where ${\cal E}_1\!=\!\int_0^\infty\!qdqV(q)[\nu\epsilon(q)\!+\!(1\!-\!\nu)\varepsilon(q)],\;$ $
{\cal E}_2\!=\!{\cal E}_{\rm SM}^{\rm HF}$ [see Eq. (1)], and ${}\;{\cal D}_{12}\!\equiv\!{\cal
D}_{21}\!=\!\sqrt{\nu}\int_0^\infty\!qdqV(q)d(q)$ with $
\epsilon\!=\!2q^2(1\!-\!q^2)^2e^{-3q^2/2}\!+\! \frac{1}{2}(4\!-\!5q^2\!+\!q^4)e^{-q^2}\!
+\!\frac{1}{16}(q^2\!-\!4)^3e^{-3q^2/4}\!+\!(2\!-{q^2}/{2})e^{-q^2/2},\;$
$\varepsilon\!=\!\frac{1}{2}[(q^2\!-\!q^4)e^{-q^2}\!-\!(q\!-\!q^3/4)^2e^{-3q^2/4}]$ {and}
$d\!=\!q^2(q^2\!-\!1)e^{-q^2}.$ [Note that expressions for $\epsilon(q)$ and $d(q)$ are the same
as in the $\nu\!=\!1$ case.] In the present experimental situation (e.g., for $B\!=\!9\,$T) we
obtain from Eq. (3) the DMA correlation shift $E={\cal E}_{\rm DM}\!=\!1.43\,$meV. The measured
and calculated magnetic field dependences of $E_{SF}$ are shown in the left-hand part of Fig.~3.
This result thus exceeds the SMA energy (1) and therefore brings the theory and the experiment
together. It is worth noting that although Eq. (3) was formally derived for any $\nu\!<\!1$, it
becomes irrelevant when gapless CDWs can be excited in the system. In that case the DMA
evidently fails and soft CDW modes coupled to the CSFE exciton should effectively reduce the
CSFE energy. Indeed, experimentally one observes a striking reduction in $E_{SF}$ when $\nu$ is
offset from the $\nu\!=\!1/3$ value corresponding to the ferromagnet state.

\begin{figure}[htb!]
\includegraphics[width=.48\textwidth]{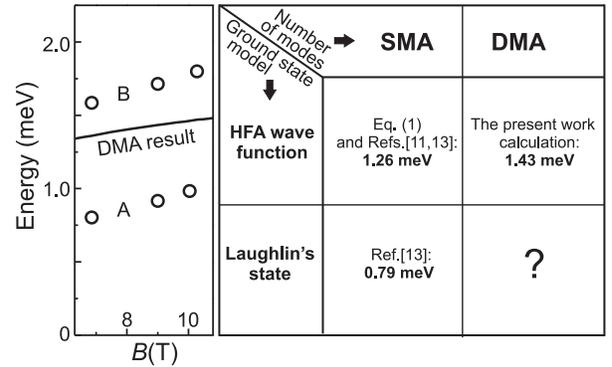}\vspace{1.7mm}
\caption{\label{fig4} Left: Experimental magnetic field dependence
of the energies for lines A and B (CSFE). The result of
calculation within DMA is also shown as a solid line. Right: Table
shows the comparison of CSFE energies calculated at $B=9$ T and
$\nu=1/3$ within existing theoretical frameworks and for
Hartree-Fock and Laughlin's ground states.}
\end{figure}

To clarify the specific place of our DMA approach among other theoretical models we present all
available results for calculations of the $\nu\!=\!1/3$ CSFE energy at $B\!=\!9\,$T in the table
(right-hand side of Fig.~3). Relevant references in this table are given in the cells located at
the crossing of the appropriate column (number of modes considered) and row (the ground state
descriptions). The cell with the question mark corresponds to the DMA where Laughlin's GS would
be used for the CSFE calculation. However, the analysis reveals that when the double-exciton
component is taken into account and the problem effectively becomes a four-particle one, the
result of the calculation (at variance with the SMA \cite{lo93,gi86}) can not be expressed in
terms of the two-particle correlation function \cite{gi84} but can only be presented in terms of
the three-particle correlation function. This function depending on three scalar arguments has
actually never been calculated and nothing is known about it at the present time.

To summarize, we have reported on the study of cyclotron spin-flip excitations in the
$\nu\!=\!1/3$ quantum Hall ferromagnet. In addition to the already known spin-flip excitation, a
new one has been observed. Its energy was measured in the range of filling factors close to
$1/3$ and was shown to ``feel'' the degree of spin polarization having a pronounced maximum at
$\nu\!=\!1/3$. At this particular filling factor the energy was compared with simulations within
the new theoretical approach alternative to the single-mode approximation. This theoretical
model takes into account the double-exciton contribution to the CSFE energy. The inclusion of
double-exciton corrections to the energy of CSFE has already substantially improved the
agreement between the experiment and the theory indicating the essentially multi-exciton nature
of CSFE in the fractional quantum Hall regime.

\vspace{2mm}

The authors would like to acknowledge support from the Russian
Foundation for Basic Research, CRDF and DFG.


\begin{thebibliography}{99}
%\bibliography{$HOME/BIB-FILES/lowD,$HOME/BIB-FILES/emp,$HOME/BIB-FILES/dots,$HOME/BIB-FILES/mikhailov}

\bibitem{Barrett1}
S.E.~Barrett, G.~Dabbagh, L.N.~Pfeiffer, K.W.~West, and R.~Tycko, Phys.~Rev.~Lett.~{\bf 74},
5112 (1995).

\bibitem{Barrett1_3}
P.~Khandelwal, N.N.~Kuzma, S.E.~Barrett, L.N.~Pfeiffer, and K.W.~West, Phys.~Rev.~Lett.~{\bf
81}, 673 (1998).

\bibitem{Larmor}
M.~Dobers, K.~von Klitzing, G.~Weimann, Phys.~Rev.~B {\bf 38},
5453 (1988).

\bibitem{nu1our}
A.B.~Van$\!$'kov, L.V.~Kulik, I.V.~Kukushkin, V.E.~Kirpichev, S.~Dickmann, V.M.~Zhilin,
J.H.~Smet, K.~v.~Klitzing, and W.~Wegscheider, Phys.~Rev.~Lett. {\bf 97}, 246801 (2006).

\bibitem{nu1temp}
A.S.~Zhuravlev, A.B.~Van$\!$'kov, L.V.~Kulik, I.V.~Kukushkin, V.E.~Kirpichev, J.H.~Smet,
K.~v.~Klitzing, V.~Umansky, and W.~Wegscheider, Phys.~Rev.~B {\bf 77}, 155404 (2008).

\bibitem{Kulik2001}
L.V.~Kulik \textit{{\it et al}.}, Phys. Rev. B {\bf 63}, 201402(R) (2001).

\bibitem{Kukushkin96}
I.V. Kukushkin, V.B. Timofeev Adv. Phys. {\bf 45}, 147 (1996).

\bibitem{2fiber}
L.V. Kulik and V.E. Kirpichev, Phys.-Usp. {\bf 49} 353 (2006) [Uspekhi Fizicheskikh Nauk {\bf
176}, 365 (2006)].

\bibitem{Donors}
A.B. Van'kov, L.V. Kulik, I.V. Kukushkin {\it et al}., JETP Lett. {\bf 87}, N3, 145 (2008).

\bibitem{di08}
S. Dickmann and V.M. Zhilin, Phys. Rev. B {\bf 78}, 115302 (2008).

\bibitem{Pinczuk1}
A.~Pinczuk, B.S.~Dennis, D.~Heiman, C.~Kallin, L.~Brey, C.~Tejedor, S.~Schmitt-Rink,
L.N.~Pfeiffer, K.W.~West, Phys.~Rev.~Lett.~{\bf 68}, 3623 (1992).

\bibitem{lu93} M. S-C. Luo, Sh.L. Chuang, S. Schmitt-Rink, and A. Pinczuk, Phys. Rev. B {\bf 48},
11086 (1993).

\bibitem{lo93} {J.P.~Longo and C.~Kallin}, Phys.~Rev.~B {\bf 47}, 4429 (1993).

\bibitem{gi86} {S.M. Girvin, A.H. MacDonald, and P.M. Platzman}, Phys. Rev. B {\bf 33}, 2481
(1986).
\bibitem{gi84} {S.M. Girvin, Phys. Rev. B {\bf 30}}, 558
(1984).


\end{thebibliography}
\end{document}